\documentclass{article}
\usepackage{spconf,amssymb,amsmath,graphicx,bm,pifont}
\usepackage{tikz,multirow,pgfplots,subcaption,setspace}
\usepackage{cite}
\title{Sandglasset: A Light Multi-Granularity Self-attentive Network For Time-Domain Speech Separation}
\name{Max W. Y. Lam$^{\star}$ \qquad Jun Wang$^{\star}$ \qquad Dan Su$^{\star}$ \qquad Dong Yu$^{\dagger}$}
\address{$^{\star}$ Tencent AI Lab, Shenzhen, China\\$^{\dagger}$ Tencent AI Lab, Bellevue WA, USA} 
\begin{document}
\ninept
\maketitle
\begin{abstract}
\small
One of the leading single-channel speech separation (SS) models is based on a TasNet with a dual-path segmentation technique, where the size of each segment remains unchanged throughout all layers. In contrast, our key finding is that multi-granularity features are essential for enhancing contextual modeling and computational efficiency. We introduce a self-attentive network with a novel sandglass-shape, namely Sandglasset, which advances the state-of-the-art (SOTA) SS performance at significantly smaller model size and computational cost. Forward along each block inside Sandglasset, the temporal granularity of the features gradually becomes coarser until reaching half of the network blocks, and then successively turns finer towards the raw signal level. We also unfold that residual connections between features with the same granularity are critical for preserving information after passing through the bottleneck layer. Experiments show our Sandglasset with only 2.3M parameters has achieved the best results on two benchmark SS datasets -- WSJ0-2mix and WSJ0-3mix, where the SI-SNRi scores have been improved by absolute 0.8 dB and 2.4 dB, respectively, comparing to the prior SOTA results.
\end{abstract}

\begin{figure*}[t]
    \centering
    \includegraphics[width=0.86\textwidth]{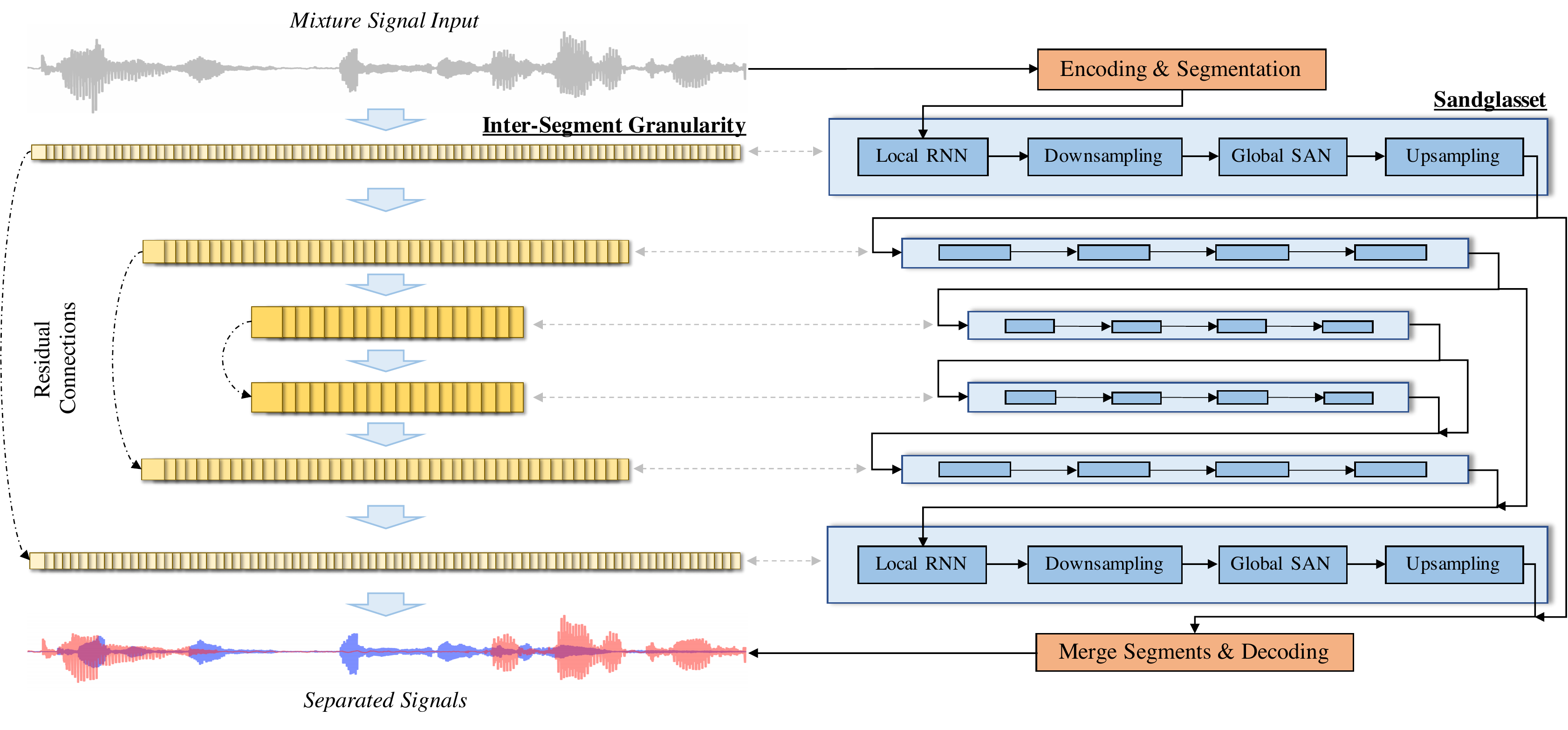}
    \caption{An illustration of the information flow inside Sandglasset. The left diagram shows the multi-granularity features with variable segment sizes that form a sandglass shape; on the right, it shows the Sandglasset blocks, each of which models a granularity depicted on the left.}
    \label{fig:1}
    \vspace{-3mm}
\end{figure*}

\section{Introduction}
\label{sec:intro}
Separating a relatively clean speech signal in the presence of multiple speaking voices is a fundamental and crucial problem (a.k.a. ``cocktail party problem'' \cite{cherry1953some,haykin2005cocktail}) for many downstream speech processing tasks \cite{narayanan2014investigation,lam2019extract,von2020end}. In this paper, we focus on the single-channel speech separation (SS) task, which is considerably more challenging than in a multi-channel setting but at the same time applies to broader scenarios, e.g., telephone conversations, many VoIP usage cases, and numerous smartphone applications. The performance of single-channel SS has been recently advanced by a variety of deep learning methods \cite{luo2018tasnet, luo2019conv, luo2019dual}. The current leading methods are based on the time-domain audio separation network (TasNet) \cite{luo2018tasnet}, which takes waveform inputs and directly reconstruct sources by computing time-domain loss with utterance-level permutation invariant training (u-PIT) \cite{yu2017permutation, kolbaek2017multitalker}. In particular, there are many variants of TasNets: the long short-term memory (LSTM) based TasNet \cite{luo2018tasnet}, the Conv-TasNet \cite{bai2018empirical, luo2019conv}, the dual-path recurrent neural network (DPRNN) \cite{luo2019dual}, the dual-path Transformer network (DPTNet) \cite{chen2020dual}, the gated DPRNN \cite{nachmani2020voice} and the Wavesplit \cite{zeghidour2020wavesplit}.
\par
Previous works of TasNets have shown that a smaller window for encoding improves the separation performance \cite{luo2019dual, chen2020dual, nachmani2020voice}, leading to much longer sequences, which poses special challenges for modeling long-term global dependencies.
To handle the very long sequences, current SOTA methods \cite{luo2019dual, chen2020dual, nachmani2020voice} employ a dual-path segmentation technique, which performs over a whole encoded sequence and divides it into intra-segment and inter-segment sequences, to which we simply refer as local and global sequences. A common strategy in prior works \cite{luo2019dual, chen2020dual, nachmani2020voice} is to use RNNs to model both the local and global sequences. Instead, we find that a self-attentive network (SAN) \cite{vaswani2017attention} would be a better structure to model the global sequence. Given an $n$-length sequence, in SAN every element can connect to another element using a direct path (i.e., in $\mathcal{O}(1)$ time) rather than recursively processing, resetting, and updating memory (i.e., in $\mathcal{O}(n)$ time) as in RNNs. Although SAN is notorious for its inefficiency in processing very long sequences due to its inherent quadratic cost, the global sequence length in the dual-path setting becomes feasible for SAN to model.
\par
Moreover, existing segmentation-based models generally use a fixed segment size unchanged throughout all layers of computation. Our finding is that the modeling capabilities of these networks could not be fully exploited if constantly modeling the global sequences with only one fixed granularity. Especially, time-domain signals essentially have different abstract contexts, e.g., phonemes, syllables, or words, at various granularity levels. Furthermore, SANs have been proven superior for modeling high-level contexts in a number of tasks \cite{shen2017disan, devlin2018bert, dai2019transformer,zhang2019self}. These together inspire us to design a new neural network architecture, where features are modeled in multi-granularity by SANs. 
Consequently, we propose a novel neural network architecture called Sandglasset, for its sandglass shape and its modest model size and complexity. Forward along each of its blocks, the granularity of the features gradually becomes coarser until reaching half of the network blocks, and then successively turns finer towards the raw signal level. We also unfold that residual connections between features with the same granularity are critical for preserving information after passing through the bottleneck layer.
\par
Finally, the proposed Sandglasset, which is very light with only 2.3M model parameters, has achieved the SOTA results on two benchmark speech separation datasets -- WSJ0-2mix and WSJ0-3mix, where the SI-SNRi scores have been pushed to 20.8 dB and 17.1 dB, surpassing the prior SOTA results by a large margin of absolute 0.8 dB and 2.4 dB. Moreover, compared to the smallest model in literature -- DPRNN, our proposed Sandglasset is remarkably lighter with 58.4\% less memory and 66.0\% fewer floating-point operations. To the best of our knowledge, Sandglasset is the first work that models multi-granularity segments using SANs in signal processing.

\section{Sandglasset}
\label{sec:2}


\subsection{Overall Architecture}
Our proposed Sandglasset is composed of $N$ blocks, as presented on the right diagram of Fig. \ref{fig:1}. If information flows from top to bottom, the first $N/2$ blocks constitute an inverted pyramid in Sandglasset, where the signal frames are successively down-sampled into shorter feature sequences of larger segments in coarser time scales, i.e., large-granularity, high-level abstract features. Then, the last $N/2$ blocks constitute a pyramid, where these high-level features are then inversely up-sampled back into longer feature sequences of smaller segments in finer time scales, i.e., fine-granularity, low-level features. To preserve information, the up-sampled features in the last $N/2$ blocks are aggregated with the earlier computed features with the same granularity using residual connections. This processing is useful for better signal reconstruction as well as avoiding gradient vanishing issues. This sandglass-shape processing strategy is capable of modeling multi-scale temporal granularity to process the input signal hierarchically and progressively, e.g., processing sounds, syllables, and words at different block levels successively. In the remainder of this section, we present the inner machinery of each module in a block as shown in the right diagram of Fig. \ref{fig:1}.

\subsection{Encoding and Segmentation}
\subsubsection{TasNet Encoder}
First of all, the input signal is a time-domain waveform mixture $\mathbf{x}\in\mathbb{R}^{T}$. Similar to other TasNet systems \cite{luo2018tasnet, luo2019conv, luo2019dual}, the input mixture signal is encoded into a sequence of 50\%-overlapping frames, denoted by $\mathbf{\Tilde{X}}=[\mathbf{\Tilde{x}}_1, ..., \mathbf{\Tilde{x}}_L]\in\mathbb{R}^{M\times L}$, where $M$ is a hyperparameter that is generally referred to as the \textit{window length}, and $L=\lceil 2T/M\rceil$. In TasNet, we use a ReLU-gated 1D convolutional layer to replace the traditional short-time Fourier transform (STFT) for signal encoding:
\begin{align}
    \mathbf{\hat{X}}=\text{ReLU}\left(\text{Conv1D}\left(\mathbf{\Tilde{X}};\mathbf{U}\right)\right),
\end{align}
where $\text{Conv1D}(\mathbf{\Tilde{X}};\mathbf{U})$ denotes the 1D convolution operation applied on $\mathbf{\Tilde{X}}$ parameterized by a learnable weight $\mathbf{U}\in \mathbb{R}^{E\times M}$ with $1\times 1$ kernels, $\text{ReLU}(\cdot)$ is the element-wise rectified linear unit used in \cite{luo2019dual, chen2020dual} to ensure non-negative outputs, and $E$ is the dimensionality of each encoded frame. Instead of directly using $\mathbf{\hat{X}}$ for the subsequent computation, we linearly map the matrix into bottleneck features $\mathbf{X}=\mathbf{B}\mathbf{\hat{X}}\in\mathbb{R}^{D\times L}$, where $\mathbf{B}\in\mathbb{R}^{D\times E}$ and $D < E$.

\subsubsection{Segmentation Module}
Given a sequence of frames in matrix form $\mathbf{X}\in\mathbb{R}^{D\times L}$, we use a segmentation module to split $\mathbf{X}$ into $S$ 50\%-overlapping segments, each of length $K$. The first and last segments are padded with zeros to create $S=\lceil 2L/K \rceil$ equal-size segments. These segments can be packed together to create a 3D tensor, denoted by $\mathcal{X}\in\mathbb{R}^{D\times K \times S}$. Note that the segment size $K$ is a hyperparameter that can be used to control the scale of the locality. The segments $\mathcal{X}$ are then passed to a stack of Sandglasset blocks.

\subsection{Sandglasset Blocks}
For the $b$-th block, we are given a 3D tensor input $\mathcal{X}_b \in \mathbb{R}^{D\times K \times S}$, enclosing $S$ segments each containing $K$ frames of $D$ dimensions. To make the following recurrence relations mathematically sound, we define $\mathcal{X}_1=\mathcal{X}$.
As shown in Fig. \ref{fig:1}, each Sandglasset block consists of mainly two operations -- firstly processing the intra-segment sequence using a recurrent neural network for modeling locality, as in \cite{luo2019dual}, and secondly modeling the inter-segment sequence using a SAN to capture the global dependencies. Interleaving with these two modules, a downsampling and an upsampling operation alter the granularity of the global sequence to be processed by the SAN.

\subsubsection{Recurrent Neural Network for Local Sequence Processing}
In our task, intra-segment sequences are the local sequences, each of length $K$, which contain subtle local details, e.g. temporal or spectral continuity, spectral structure, timbre, etc., which are rather irrelevant to the long-term context.
In Sandglasset, we assign the local sequence processing task to a one-layer RNN. Specifically, in each Sandglasset block, the 3D tensor $\mathcal{X}^{LR}_b=\mathcal{X}_b$ obtained from the segmentation process is passed to a bi-directional LSTM of $H$ hidden nodes. Here, for ease of reference, we use $\mathcal{X}^{LR}$ and $\mathcal{Y}^{LR}$ to respectively denote the inputs for the local RNN and the outputs from the local RNN. The superscript $LR$ is used to differentiate from the corresponding input-output pairs in the global SAN model.
\begin{align}
    \mathcal{Y}^{LR}_b = \left[\mathbf{M}_b\cdot \text{BiLSTM}_{b}\left(\mathcal{X}^{LR}_b[:, s, :]\right)+\mathbf{c}_b, s = 1, . . . , S\right],
\end{align}
where $\cdot$ is used to denote matrix multiplication, $\mathcal{X}^{LR}_b[:, s, :]\in \mathbb{R}^{D\times K}$ refers to the local sequence within the $s$-th chunk, $\mathbf{M}_b\in \mathbb{R}^{D \times 2H}$ and $\mathbf{c}_b\in \mathbb{R}^{D}$ are the parameters of a linear transformation.

\subsubsection{Self-Attentive Network for Multi-Granularity Modeling}
After processing the intra-segment sequences each of length $K$, we aim at modeling the inter-segment sequences, each of length $S$. Noted that inter-segment sequences are likely to encode the contextual information of the speech signal. In Sandglasset, we employ a variable-context-aware self-attentive network (SAN) to capture the global dependencies in different time scales. 
\par
Instead of directly taking $\mathcal{Y}^{LR}_b$ as the input to a SAN, we first apply a layer normalization operation $\text{LN}(\cdot)$ to the LR layer's output and add a residual connection to the block input:
\begin{align}
    \mathcal{X}^{GA}_b&=\text{LN}\left(\mathcal{Y}^{LR}_b\right)+\mathcal{X}_b,
\end{align}
which is then re-sampled to modify the time scale for global processing across segments:
\begin{align}
    \mathcal{Y}^{GA}_b &= \text{US}_b\left(\text{SAN}_b\left(\text{DS}_b\left(\mathcal{X}^{GA}_b\right)\right)\right)
\end{align}
where $\text{US}_b(\cdot)$ and $\text{DS}_b(\cdot)$ are the upsampling and downsampling operations, respectively, which are defined as the follows:
\begin{align}
\label{eq:us}
\text{US}_b\left(\mathcal{X}\right)&=
\begin{cases} 
\text{ConvTrans1D}_K\left(\mathcal{X};4^{b}\right) & \text{if}\,\,\,\,\,\, b \leq N/2; \\
\text{ConvTrans1D}_{K}\left(\mathcal{X};4^{N-b-1}\right) & \text{if}\,\,\,\,\,\, b > N/2,
\end{cases}\\
\label{eq:ds}
\text{DS}_b\left(\mathcal{X}\right)&=
\begin{cases} 
\text{Conv1D}_{K}\left(\mathcal{X}; 4^{b}\right) & \text{if}\,\,\,\,\,\, b \leq N/2; \\
\text{Conv1D}_{K}\left(\mathcal{X}; 4^{N-b-1}\right) & \text{if}\,\,\,\,\,\, b > N/2,
\end{cases}
\end{align}
where $\text{Conv1D}_{A}(\cdot;B)$ and $\text{ConvTran1D}_{A}(\cdot;B)$ respectively denote the 1D and 1D transposed convolution operations along the axis of length $A$ with a kernel size of $B$ and a stride length of $B$ such that the resultant length becomes $\lfloor A/B\rfloor$ (in DS) or $\lfloor AB\rfloor$ (in US) long.
We also employ the variable-context-aware self-attentive network $\text{SAN}_b(\cdot)$, which is modified from the pioneering work \cite{vaswani2017attention}. For simplicity, we generally define our SAN for any input $\mathcal{X}\in\mathbb{R}^{D\times S \times K}$:
\begin{align}
    \text{SAN}(\mathcal{X}) = \left[\text{SelfAttn}\left(\text{LN}\left(\mathcal{X}[:, :, k]\right)+\mathbf{P}\right), k = 1, . . . , K\right],
\end{align}
where $\mathbf{P}$ denotes the positional encoding matrix as introduced in \cite{vaswani2017attention}, and $\mathcal{X}[:, :, k]\in \mathbb{R}^{D\times S}$ refers to the inter-segment sequence. Here, $\text{SelfAttn}(\cdot)$ is a typical multi-head self-attention function that linearly projects an input matrix $\mathbf{X}\in\mathbb{R}^{D\times S}$ into three forms of matrices, commonly denoted as query $\mathbf{Q}_j$, key $\mathbf{K}_j$, and value $\mathbf{V}_j$ matrices to compute the scaled dot-product attention for different heads $j=1,...,J$, which are finally combined by a concatenation plus a matrix multiplication:
\begin{align}
\left[\mathbf{Q}_j\,\,\, \mathbf{K}_j \,\,\,\mathbf{A}_j \right]^\top&= \left[\mathbf{W}^\text{Q}_j\,\,\, \mathbf{W}^\text{K}_j \,\,\,\mathbf{W}^\text{V}_j \right]^\top\mathbf{X}+\left[\mathbf{b}^\text{Q}_j\,\,\, \mathbf{b}^\text{K}_j \,\,\,\mathbf{b}^\text{V}_j \right]^\top\\
\mathbf{A}_j &= \text{Softmax}\left(\frac{\mathbf{Q}_j^\top\mathbf{K}_j}{\sqrt{D/J}}\right)\mathbf{V}_j\\
\mathbf{A}&= \mathbf{W}\cdot\text{Concat}\left(\mathbf{A}_{1},...,\mathbf{A}_{J}\right)\\
\text{SelfAttn}(\mathbf{X})&=\text{LN}(\mathbf{X}+\text{DROP}(\mathbf{A}))
\end{align}
where $\text{DROP}(\cdot)$ denotes the dropout technique \cite{srivastava2014dropout}, and $\mathbf{W}\in\mathbb{R}^{D\times D}$, $\mathbf{W}^\text{Q}_j, \mathbf{W}^\text{K}_j, \mathbf{W}^\text{V}_j \in \mathbb{R}^{D/J \times D}$ and $\mathbf{b}^\text{Q}_j, \mathbf{b}^\text{K}_j, \mathbf{b}^\text{V}_j \in \mathbb{R}^{D/J}$ are the parameters for SAN.

\subsubsection{Residual Connections to Prevent Information Loss}
One of the highlights in Sandglasset is to add residual connections between pairs of Sandglasset blocks that are of the same granularity. This technique is used to prevent information loss after passing through the middle blocks, where the granularity is on the coarsest scale. Mathematically, we define
\begin{align}
\mathcal{X}^{LR}_{b+1} &=
\begin{cases}
\mathcal{Y}^{GA}_b & \text{if}\,\,\,\,\,\, b\leq N/2;\\
\mathcal{Y}^{GA}_b + \mathcal{Y}^{GA}_{b-N/2}& \text{if}\,\,\,\,\,\, b > N/2,
\end{cases} 
\end{align}
which also defines the recurrence relation between the $b$-th and the $(b+1)$-th Sandglasset block. Our experimental result indicates that in practice adding residual connections is critical to remedy raw signal level details for improving signal reconstruction and to avoid gradient vanishing issues for better parameter learning. 

A seminal work in signal processing -- U-Net \cite{ronneberger2015u, daniel2018unet, choi2020phase} seems a similar idea to ours for re-sampling and combining features at different time scales. Nonetheless, Sandglasset is very different in many aspects: (1) we have downsampling and upsampling operations together performed in one block; (2) our multi-granularity features are only processed by the SANs within each block; and (3) the residual connections across Sandglasset blocks are purely based on addition.
\par
\subsection{Merge Segments and Decoding}
\subsubsection{Mask Estimation}
After passing through $N$ Sandglass blocks, we obtain a 3D tensor output $\mathcal{X}_{N+1}^{LR}\in\mathbb{R}^{D\times K\times S}$, which can be used to estimate masks for $C$ sources. To do so, we first transform the last block's output using a PReLU-gated 2D convolutional layer to obtain a 4D tensor of shape $C\times E \times K \times S$:
\begin{align}
    \mathcal{Y}=\text{Conv2D}\left(\text{PReLU}\left(\mathcal{X}_{N+1}^{LR}\right);\mathbf{C}\right),
\end{align}
where $\text{Conv2D}(\mathcal{Y};\mathbf{C})$ denotes the 2D convolution operation applied on $\mathcal{Y}$ parameterized by a learnable weight $\mathbf{C}\in \mathbb{R}^{CE\times D}$ with a $1\times 1$ kernel, $\text{PReLU}(\cdot)$ is the element-wise parametric ReLU. We then merge the output segments $\mathcal{Y}$ using an $\text{OverlapAdd}$\footnote{https://github.com/tensorflow/tensorflow/blob/r1.12/tensorflow/contrib/\\signal/python/ops/reconstruction\_ops.py} approach \cite{luo2019dual} to match the shape of the mixture frames $\mathbf{\hat{X}}\in\mathbb{R}^{E\times L}$ for masking:
\begin{align}
    \mathbf{M} &=\text{ReLU}\left(\text{OverlapAdd}\left(\mathcal{Y}\right)\right),
\end{align}
where $\odot$ is the element-wise product operation.

\subsubsection{Decoder for Waveform Reconstruction}
Finally, the $c$-th source signal is reconstructed by applying the $c$-th estimated mask to the initially computed mixture frames $\mathbf{\hat{X}}$ and then using OverlapAdd to merge frames into waveform:
\begin{align}
    \hat{\mathbf{s}}_c = \text{OverlapAdd}(\Tilde{\mathbf{X}}\odot \mathbf{M}_c).
\end{align}
Last but not least, given $C$ estimated sources, the scale-invariant source-to-noise ratio (SI-SNR) loss \cite{luo2018tasnet} is used with u-PIT \cite{yu2017permutation} to learn the network parameters and to solve the permutation problem.

\section{Experiments}
\label{sec:4}
\subsection{Experimental Setup}
\subsubsection{Data}
To compare with the SOTA speech separation networks, we used two benchmark datasets for evaluation -- WSJ0-2mix and WSJ0-3mix \cite{hershey2016deep}, which are generated from the Wall Street
Journal (WSJ0) \cite{garofalocontinuous} dataset by randomly mixing clean utterances from different speakers at a sampling rate of 8 kHz with SNRs between 0 dB and 5 dB. The separation datasets consist of 30 hours of training, 10 hours of validation, and 5 hours of test data from 16 unseen speakers. Both WSJ0-2mix and WSJ0-3mix have been widely used as the benchmark in single-channel speech separation \cite{luo2019dual, liu2019divide, luo2018tasnet, luo2019conv, wang2018end, zhang2020furcanext, lam2020mixup}. 

\subsubsection{Implementation Details}
In our implementation, we used the setting of encoder-decoder modules in \cite{luo2018tasnet, luo2019conv} and the segmentation module described in \cite{luo2019dual}. In particular, we set $M=4$, $E=256$, and $D=128$. For Sandglasset, we used 6 Sandglasset blocks, i.e., $N=6$. In the first Sandglasset block, we used an initial segment size $K=256$, which would be shortened/prolonged by a factor of 4 in the first/last three blocks, as described in Eq. (\ref{eq:us}-\ref{eq:ds}). Within each Sandglasset block, we used local Bi-LSTM with 128 hidden units, i.e., $H=128$. The global SAN was set to be 8-head, i.e., $J=8$ with a 0.1 dropout rate. For training, we used Adam \cite{kingma2014adam} optimizer with an initial learning rate of $0.001$ and a decaying rate of $0.98$. The optimization was stopped if no lower validation loss was obtained for $10$ consecutive epochs.

\subsubsection{Mixing Same-Speaker Utterances as Post Training}
\label{sec:3.1.3}
By inspecting the poor separation cases of Sandglasset in WSJ0-2mix, we found that those mixture inputs shared a common characteristic that both speakers have a similar voice timbre, so that the model may keep both voices in the two output signals. This reveals that our model's separation is highly dependent on the voice timbre of different speakers. We conceived that one of the main reasons is that mixture inputs with similar voice timbres are rare in the training set, which makes it hard for our model to learn to differentiate those similar voices. To alleviate this problem, we designed a simple, easy-to-implement post training method for Sandglasset. In particular, after the convergence of the normal training, we expanded the training set by adding dynamically mixed utterances from the same speaker in a 1:1 ratio in sample size relative to the original training data.

\subsection{Performance Comparisons}
\begin{table}[t]
\caption{Comparison of performances on the WSJ0-2mix test set. The models that exploit speaker IDs as additional information for training and testing are marked with ``+ Spk ID''. $^{\dagger}$ denotes our estimated model size based on the authors' description. }
\label{tab:1}
\centering

\begin{tabular}{c|c|c|c}
\hline
\textbf{Model} & \textbf{Params.} & \textbf{SI-SNRi} & \textbf{SDRi} \\
\hline
{BLSTM-TasNet \cite{luo2018tasnet}} & 23.6M & 13.2 & 13.6 \\
{Conv-TasNet \cite{luo2019conv}} & 8.8M & 15.3 & 15.6 \\
{Conv-TasNet + MBT \cite{lam2020mixup}} & 8.8M & 15.5 & 15.9 \\
{FurcaNeXt \cite{zhang2020furcanext}} & 51.4M & 18.4 & - \\
{DPRNN \cite{luo2019dual}} & 2.6M & 18.8 & 19.1 \\
{DPTNet \cite{chen2020dual}} & 2.7M & 20.2 & 20.6 \\
{\bf{Sandglasset (w/o RES)}} & \bf{2.3M} & 20.1 & 20.3 \\
{\bf{Sandglasset (SG)}} & \bf{2.3M} & 20.3 & 20.5 \\
{\bf{Sandglasset (MG)}} & \bf{2.3M} & 20.8 & 21.0 \\
{\bf{Sandglasset (MG) + PT}} & \bf{2.3M} & \bf{21.0} & \bf{21.2} \\ \hline
{Gated DPRNN + Spk ID \cite{nachmani2020voice}} & 7.5M & 20.1 & - \\
{Wavesplit + Spk ID \cite{zeghidour2020wavesplit}} & $^{\dagger}$42.5M & \bf{21.0} & \bf{21.2} \\
\hline
\end{tabular}
\end{table}

\begin{table}[t]
\caption{Comparison of performances on the WSJ0-3mix test set.}
\label{tab:2}
\centering

\begin{tabular}{c|c|c|c}
\hline
\textbf{Model} & \textbf{Params.} & \textbf{SI-SNRi} & \textbf{SDRi} \\
\hline
{Conv-TasNet \cite{luo2019conv}} & 8.8M & 12.7 & 13.1 \\
{DPRNN \cite{luo2019dual}} & 2.6M & 14.7 & -  \\ 
{\bf{Sandglasset (MG)}} & \bf{2.3M} & \bf{17.1} & \bf{17.4} \\\hline
{Gated DPRNN + Spk ID \cite{nachmani2020voice}} & 7.5M & 16.7 & - \\
{Wavesplit + Spk ID \cite{zeghidour2020wavesplit}} & $^{\dagger}$42.5M & \bf{17.3} & \bf{17.6} \\
\hline
\end{tabular}
\vspace{-3mm}
\end{table}
The SISNRi and SDRi performances of Sandglasset in WSJ0-2mix are reported in Table \ref{tab:1}. First of all, for an ablation study on our proposed multi-granularity (MG) strategy, we trained an ablated baseline system -- ``Sandglasset (SG)", in which each Sandglasset block uses a single-granularity strategy with a fixed segment size ($K=256$). Comparing ``Sandglasset (SG)" to ``Sandglasset (MG)", we can see a significant drop in SI-SNRi and SDRi scores if Sandglasset was deprived of the multi-granularity mechanism. This asserts our initial expectation that multi-granularity can better exploit SANs for modeling multi-level contexts. For another ablation study, we trained a Sandglasset without residual connections, denoted by ``Sandglasset (w/o RES)", which produced a much-degraded performance. We also found that by using the simple post training strategy described in Section \ref{sec:3.1.3}, the performance of Sandglasset can be further improved.

Overall, the proposed Sandglasset has achieved the best separation performance with parameters as few as 2.3M, which is the lightest model size that is ever reported for the SS tasks. We would like to emphasize that, to focus on studying the advantage of the network architecture only, we purposely avoid using any speaker information to help further increase the scores of Sandglasset, unlike what has been done in the two most recent systems ``Gated DPRNN + Spk ID"  \cite{nachmani2020voice} and ``Wavesplit + Spk ID" \cite{zeghidour2020wavesplit}. Comparing to the strongest reference model regardless of speaker information, Sandglasset has attained an absolute improvement of 0.8 dB SI-SNRi. The WSJ0-3mix result of Sandglasset, as shown in Table \ref{tab:2}, also consistently shows an absolute improvement of 2.4 dB SI-SNRi over the best reference model with no speaker information. 

\subsection{Computational Cost Analysis}
\label{sec:cost}
Moreover, thanks to some coarser-scale global processing, another merit of Sandglasset is a significant reduction in computational cost, relative to a model that is comparable in size -- DPRNN. In Table \ref{tab:3}, we reported the runtime memory and the floating-point operations (FLOPs) \footnote{https://github.com/sovrasov/flops-counter.pytorch} which indicates the model efficiency for processing each second of mixture input. Finally, compared to the best performing DPRNN (i.e., 2-sample window), Sandglasset consumed 58.4\% less memory and 66.0\% fewer FLOPs.
\begin{table}[t]
\caption{Comparison of computational costs.}
\label{tab:3}
\centering
\begin{tabular}{c|c|c|c}
\hline
\textbf{Model} & \textbf{Params.} & \textbf{Memory} (GB) & \textbf{GFLOPs} ($10^{9}$) \\
\hline
{DPRNN \cite{luo2019dual}} & 2.6M & 1.97 & 84.7  \\ 
{\bf{Sandglasset}} & \bf{2.3M} & \bf{0.82} ($\downarrow$58.4\%)  & \bf{28.8} ($\downarrow$66.0\%)  \\
\hline
\end{tabular}
\vspace{-3mm}
\end{table}
\section{Conclusions}
\label{sec:conc}
This paper proposes a novel sandglass-shape network for time-domain single-channel speech separation, namely Sandglasset. This advanced network architecture combines the advantages of the self-attention networks and the proposed multi-granularity mechanism to hierarchically and progressively model high-level, large-granularity contexts and low-level, fine-granularity details. In our experiment, Sandglasset achieved state-of-the-art results on two benchmark datasets, especially, with the lightest model size that has ever been reported for SS tasks. Comparing to the previous smallest and strongest model in literature, our proposed model is also very light in terms of memory (58.4\% less) and computations (66\% fewer), which suggests Sandglasset a more economical and practical model for industrial deployment.


\bibliographystyle{IEEEbib}
\bibliography{strings,refs}

\begin{thebibliography}{10}

\bibitem{cherry1953some}
E~Colin Cherry,
\newblock ``Some experiments on the recognition of speech, with one and with
  two ears,''
\newblock {\em The Journal of the acoustical society of America}, vol. 25, no.
  5, pp. 975--979, 1953.

\bibitem{haykin2005cocktail}
Simon Haykin and Zhe Chen,
\newblock ``The cocktail party problem,''
\newblock {\em Neural computation}, vol. 17, no. 9, pp. 1875--1902, 2005.

\bibitem{narayanan2014investigation}
Arun Narayanan and DeLiang Wang,
\newblock ``Investigation of speech separation as a front-end for noise robust
  speech recognition,''
\newblock {\em IEEE/ACM Transactions on Audio, Speech, and Language
  Processing}, vol. 22, no. 4, pp. 826--835, 2014.

\bibitem{lam2019extract}
Max~WY Lam, Jun Wang, Xunying Liu, Helen Meng, Dan Su, and Dong Yu,
\newblock ``Extract, adapt and recognize: an end-to-end neural network for
  corrupted monaural speech recognition,''
\newblock {\em Proc. INTERSPEECH}, pp. 2778--2782, 2019.

\bibitem{von2020end}
Thilo von Neumann, Keisuke Kinoshita, Lukas Drude, Christoph Boeddeker, Marc
  Delcroix, Tomohiro Nakatani, and Reinhold Haeb-Umbach,
\newblock ``End-to-end training of time domain audio separation and
  recognition,''
\newblock in {\em ICASSP 2020-2020 IEEE International Conference on Acoustics,
  Speech and Signal Processing (ICASSP)}. IEEE, 2020, pp. 7004--7008.

\bibitem{luo2018tasnet}
Yi~Luo and Nima Mesgarani,
\newblock ``Tasnet: time-domain audio separation network for real-time,
  single-channel speech separation,''
\newblock in {\em Proc. ICASSP}. IEEE, 2018, pp. 696--700.

\bibitem{luo2019conv}
Yi~Luo and Nima Mesgarani,
\newblock ``Conv-tasnet: Surpassing ideal time--frequency magnitude masking for
  speech separation,''
\newblock {\em IEEE/ACM transactions on audio, speech, and language
  processing}, vol. 27, no. 8, pp. 1256--1266, 2019.

\bibitem{luo2019dual}
Yi~Luo, Zhuo Chen, and Takuya Yoshioka,
\newblock ``Dual-path rnn: efficient long sequence modeling for time-domain
  single-channel speech separation,''
\newblock {\em arXiv preprint arXiv:1910.06379}, 2019.

\bibitem{yu2017permutation}
Dong Yu, Morten Kolb{\ae}k, Zheng-Hua Tan, and Jesper Jensen,
\newblock ``Permutation invariant training of deep models for
  speaker-independent multi-talker speech separation,''
\newblock in {\em Proc. ICASSP}. IEEE, 2017, pp. 241--245.

\bibitem{kolbaek2017multitalker}
Morten Kolb{\ae}k, Dong Yu, Zheng-Hua Tan, Jesper Jensen, Morten Kolbaek, Dong
  Yu, Zheng-Hua Tan, and Jesper Jensen,
\newblock ``Multitalker speech separation with utterance-level permutation
  invariant training of deep recurrent neural networks,''
\newblock {\em TASLP}, vol. 25, no. 10, pp. 1901--1913, 2017.

\bibitem{bai2018empirical}
Shaojie Bai, J~Zico Kolter, and Vladlen Koltun,
\newblock ``An empirical evaluation of generic convolutional and recurrent
  networks for sequence modeling,''
\newblock {\em arXiv preprint arXiv:1803.01271}, 2018.

\bibitem{chen2020dual}
Jingjing Chen, Qirong Mao, and Dong Liu,
\newblock ``Dual-path transformer network: Direct context-aware modeling for
  end-to-end monaural speech separation,''
\newblock {\em arXiv preprint arXiv:2007.13975}, 2020.

\bibitem{nachmani2020voice}
Eliya Nachmani, Yossi Adi, and Lior Wolf,
\newblock ``Voice separation with an unknown number of multiple speakers,''
\newblock {\em arXiv preprint arXiv:2003.01531}, 2020.

\bibitem{zeghidour2020wavesplit}
Neil Zeghidour and David Grangier,
\newblock ``Wavesplit: End-to-end speech separation by speaker clustering,''
\newblock {\em arXiv preprint arXiv:2002.08933v1}, 2020.

\bibitem{vaswani2017attention}
Ashish Vaswani, Noam Shazeer, Niki Parmar, Jakob Uszkoreit, Llion Jones,
  Aidan~N Gomez, {\L}ukasz Kaiser, and Illia Polosukhin,
\newblock ``Attention is all you need,''
\newblock in {\em Advances in neural information processing systems}, 2017, pp.
  5998--6008.

\bibitem{shen2017disan}
Tao Shen, Tianyi Zhou, Guodong Long, Jing Jiang, Shirui Pan, and Chengqi Zhang,
\newblock ``Disan: Directional self-attention network for rnn/cnn-free language
  understanding,''
\newblock {\em arXiv preprint arXiv:1709.04696}, 2017.

\bibitem{devlin2018bert}
Jacob Devlin, Ming-Wei Chang, Kenton Lee, and Kristina Toutanova,
\newblock ``Bert: Pre-training of deep bidirectional transformers for language
  understanding,''
\newblock {\em arXiv preprint arXiv:1810.04805}, 2018.

\bibitem{dai2019transformer}
Zihang Dai, Zhilin Yang, Yiming Yang, Jaime Carbonell, Quoc~V Le, and Ruslan
  Salakhutdinov,
\newblock ``Transformer-xl: Attentive language models beyond a fixed-length
  context,''
\newblock {\em arXiv preprint arXiv:1901.02860}, 2019.

\bibitem{zhang2019self}
Han Zhang, Ian Goodfellow, Dimitris Metaxas, and Augustus Odena,
\newblock ``Self-attention generative adversarial networks,''
\newblock in {\em International Conference on Machine Learning}. PMLR, 2019,
  pp. 7354--7363.

\bibitem{srivastava2014dropout}
Nitish Srivastava, Geoffrey Hinton, Alex Krizhevsky, Ilya Sutskever, and Ruslan
  Salakhutdinov,
\newblock ``Dropout: a simple way to prevent neural networks from
  overfitting,''
\newblock {\em The journal of machine learning research}, vol. 15, no. 1, pp.
  1929--1958, 2014.

\bibitem{ronneberger2015u}
Olaf Ronneberger, Philipp Fischer, and Thomas Brox,
\newblock ``U-net: Convolutional networks for biomedical image segmentation,''
\newblock in {\em International Conference on Medical image computing and
  computer-assisted intervention}. Springer, 2015, pp. 234--241.

\bibitem{daniel2018unet}
Daniel Stoller, Sebastian Ewert, and Simon Dixon,
\newblock ``Wave-u-net: A multi-scale neural network for end-to-end audio
  source separation,''
\newblock in {\em 19th International Society for Music Information Retrieval
  Conference, ISMIR}, 2018.

\bibitem{choi2020phase}
Hyeong-Seok Choi, Hoon Heo, Jie~Hwan Lee, and Kyogu Lee,
\newblock ``Phase-aware single-stage speech denoising and dereverberation with
  u-net,''
\newblock {\em arXiv preprint arXiv:2006.00687}, 2020.

\bibitem{hershey2016deep}
John~R Hershey, Zhuo Chen, Jonathan Le~Roux, and Shinji Watanabe,
\newblock ``Deep clustering: Discriminative embeddings for segmentation and
  separation,''
\newblock in {\em Proc. ICASSP}. IEEE, 2016, pp. 31--35.

\bibitem{garofalocontinuous}
J~Garofalo, D~David~Graff, D~Paul, and D~Pallett,
\newblock ``Continuous speech recognition (csr-i) wall street journal (wsj0)
  news, complete. linguistic data consortium, philadelphia (1993),'' .

\bibitem{liu2019divide}
Yuzhou Liu and DeLiang Wang,
\newblock ``Divide and conquer: A deep casa approach to talker-independent
  monaural speaker separation,''
\newblock {\em IEEE/ACM Transactions on Audio, Speech, and Language
  Processing}, vol. 27, no. 12, pp. 2092--2102, 2019.

\bibitem{wang2018end}
Zhong-Qiu Wang, Jonathan~Le Roux, DeLiang Wang, and John~R Hershey,
\newblock ``End-to-end speech separation with unfolded iterative phase
  reconstruction,''
\newblock {\em arXiv preprint arXiv:1804.10204}, 2018.

\bibitem{zhang2020furcanext}
Liwen Zhang, Ziqiang Shi, Jiqing Han, Anyan Shi, and Ding Ma,
\newblock ``Furcanext: End-to-end monaural speech separation with dynamic gated
  dilated temporal convolutional networks,''
\newblock in {\em International Conference on Multimedia Modeling}. Springer,
  2020, pp. 653--665.

\bibitem{lam2020mixup}
Max~WY Lam, Jun Wang, Dan Su, and Dong Yu,
\newblock ``Mixup-breakdown: a consistency training method for improving
  generalization of speech separation models,''
\newblock {\em Proc. ICASSP}, 2020.

\bibitem{kingma2014adam}
Diederik~P Kingma and Jimmy Ba,
\newblock ``Adam: A method for stochastic optimization,''
\newblock {\em arXiv preprint arXiv:1412.6980}, 2014.

\end{thebibliography}

\end{document}